\documentclass[12pt]{article}
\usepackage{epsf}
\usepackage{latexsym}
\newcommand{\be}{\begin{equation}}
\newcommand{\ee}{\end{equation}}
\newcommand{\ba}{\begin{array}}
\newcommand{\ea}{\end{array}}
\newcommand{\bqa}{\begin{eqnarray}}
\newcommand{\eqa}{\end{eqnarray}}
\begin{document}

\begin{center}
{\Large\bf The [1,2] Pad\'e Amplitudes for $\pi\pi$ Scatterings in
 Chiral Perturbation Theory}
\\[10mm]
{\sc  Guang-You Qin, W.~Z.~Deng, Z.~G.~Xiao and H.~Q.~Zheng}
\\[5mm]
{\it Department of Physics, Peking University, Beijing 100871,
P.~R.~China }
\\[5mm]
\begin{abstract}
A detailed analysis to the [1,2] Pad\'e approximation to the
$\pi\pi$ scattering 2--loop amplitudes in  chiral perturbation
theory is made.
\end{abstract}
\end{center}
\vspace{1cm}

The chiral perturbation theory (ChPT)
(\cite{Weinberg01}--\cite{Gasser02}) is a powerful tool in
studying strong interaction physics at low energies and has been
extensively studied at 1--loop level~\cite{review}. The 2--loop
results are also available in recent years
(\cite{2loops1}--\cite{2loops}). However, since the chiral
expansion is an expansion in terms of the external momentum, the
perturbation series to any finite order diverges very rapidly at
high energies. Therefore the violation of unitarity gets even
worse for 2--loop amplitudes than the 1--loop amplitudes at high
energies. Also the number of parameters in the effective
Lagrangian which are not fixed by symmetry alone increases
rapidly. Therefore increasing the order of the perturbation
expansion does not work at all  for the purpose of exploring
physics in the non-perturbative region, or at higher energies and
non-perturbative studies become necessary. A widely used method to
remedy the violation of unitarity is the so called Pad\'e
approximation.~\footnote{Variations of the Pad\'e approximation
method can be found in the literature named as the inverse
amplitude method (IAM)~\cite{IAM} and the chiral unitarization
approach~\cite{pade} with somewhat different formalism and
motivation.} A nice feature of the Pad\'e approximation is that it
restores unitarity with full respect,  at low energies, to the
available information from perturbation theory. Therefore, even
though it is well known that it violates crossing symmetry, Pad\'e
approximation is considered to be a valuable tool in exploring
physics in the non-perturbative region, such as the properties of
physical resonances. However, a previous study~\cite{ang}
indicates that the [1,1] Pad\'e approximation encounters a serious
problem by predicting spurious physical sheet resonances (SPSRs).
Usually these SPSRs locate at distant places very far from the
region where the perturbation results are valid. The predictions
of the Pad\'e approximants constructed from the perturbative
amplitudes should, of course not,  be considered as meaningful in
the region far away from the region where perturbation theory
remains to be valid. One may further argue that since those SPSRs
are far from the region we are concerning the use of the Pad\'e
approximation is still acceptable at least in  phenomenological
discussions. However, the problem with Pad\'e approximation is not
only because it predicts SPSRs in the distant region too far away
to be worthwhile to pay any attention, but also because  those
SPSRs usually have large couplings to $\pi\pi$ which lead to
strong influence to the region we are really interested in, and
hence their existence casts doubt on the remaining predictions of
the Pad\'e approximants which might otherwise  be assumed as
meaningful, at least at quantitative level. The aim of the present
study is to further investigate the Pad\'e approximation following
the method of Ref.~\cite{ang}. We will extend the work of
Ref.~\cite{ang} by also analyzing the [1,2] Pad\'e approximants,
since the 2--loop perturbation results are already available. One
of the main motivation of the present work is to investigate the
possibility that the [1,2] Pad\'e approximants can rescue, to some
extent, the bad situation the [1,1] Pad\'e approximants encounter.
The conclusion is rather $negative$, as we will see in the
following text. However, we believe it is still worthwhile to
present our results. Since the Pad\'e approximation is a very
popular approximation method widely used in phenomenological
discussions, we hope the presentation of the present work could
benefit physicists who are working in the related fields of
non-perturbative physics.

For the $\pi\pi\rightarrow\pi\pi$ scattering, it is well known
that the isospin
amplitudes in the $s$ channel can be decomposed as,%
\bqa%
T^{I=0}(s,t,u) &=& 3A(s,t,u)+A(t,u,s)+A(u,s,t)\ ,\nonumber \\%
T^{I=1}(s,t,u) &=& A(t,u,s)-A(u,s,t)\ ,\nonumber \\%
T^{I=2}(s,t,u) &=& A(t,u,s)+A(u,s,t)\ ,
\eqa%
where $s,t,u$ are the usual Mandelstam variables,
\bqa%
{s={(p_1+p_2)^2/{M_\pi^2}}}\ ,\ \ %
{t={(p_3-p_1)^2/{M_\pi^2}}}\ ,\ \ %
{u={(p_4-p_1)^2/{M_\pi^2}}}\ .%
\eqa%
In $SU(2)\times SU(2)$ chiral perturbation theory to two
loops~\cite{2loops},\ the momentum expansion of the
amplitudes amounts to a Taylor series in%
\be%
x_2 = {M_\pi^2 \over F_\pi^2}\ .\nonumber%
\ee%
The amplitudes can be expressed in terms of six parameters $b_1$,\
$b_2$,\ $b_3$,\ $b_4$,\ $b_5$,\ $b_6$,
\bqa\label{A(s,t,u)}%
A(s,t,u)&=& x_2(s-1)\nonumber \\%
&&+x_2^2\left(b_1+b_2s+b_3s^2+b_4(t-u)^2\right)\nonumber \\%
&&+x_2^2\left(F^{(1)}(s)+G^{(1)}(s,t)+G^{(1)}(s,u)\right)\nonumber \\%
&&+x_2^3\left(b_5s^3+b_6s(t-u)^2\right)\nonumber \\%
&&+x_2^3\left(F^{(2)}(s)+G^{(2)}(s,t)+G^{(2)}(s,u)\right)\nonumber \\%
&&+O(x_2^4)\ .%
\eqa%
The expressions of the functions $F^{(i)}(s)$ and $G^{(i)}(s,t)$
and the constants
$b_i$ are rather lengthy and we refer to the original
work~\cite{2loops} for the details.%

The partial wave expansion of the isospin amplitudes is written
as%
\be%
T^I(s,t,u) = 32\pi \sum_J(2J+1)P_J(\cos\theta){T^I_J}(s)\ .%
\ee%
The inverse expression is%
\bqa%
T^I_J(s) &=& {1\over 64\pi} \int_{-1}^1
d({\cos\theta}) P_J(cos\theta)T^{I}(s,t,u)\ ,\nonumber \\%
cos\theta &=& 1+{{2t}\over {s-4}}\ ,\nonumber \\%
u &=& 4-s-t\ .%
\eqa%
The partial wave amplitude in ChPT  expanded to $O(p^6)$ is,%
\be%
T^I_J(s) = T^I_{J, 2}(s)+T^I_{J,4}(s)+T^I_{J,6}(s)\ .%
\ee%
In Ref.~\cite{ang} we have discussed the [1,1] Pad\'e approximants
of the partial wave amplitudes in 1-loop ChPT.\ To proceed we now
construct the [1,2] Pad\'e approximants to the partial wave
amplitudes in 2--loop ChPT,
\be\label{pade}%
{T^I_J}^{[1,2]}(s) = { T^I_{J,2}(s)\over { 1 - {T^I_{J,4}(s)\over
T^I_{J,2}(s)} - {T^I_{J,6}(s)\over T^I_{J,2}(s)} + \left({
T^I_{J,4}(s)\over T^I_{J,2}(s) }\right)^2 } }\ .
\ee%
Perturbation theory satisfies the elastic unitarity relation,%
\be\label{unitarity}%
{\rm Im}{T^I_J(s)} = \rho(s) |T^I_J(s)|^2\ ,%
\ee%
at each order of the perturbation expansion
in powers of the quark masses and external momentum, i.e.,%
\bqa%
{\rm Im}{T^I_{J,2}}(s) &=& 0\ ,\nonumber \\%
{\rm Im}{T^I_{J,4}}(s) &=& \rho(s) \left(T^I_{J,2}(s)\right)^2\ ,\nonumber \\%
{\rm Im}{T^I_{J,6}}(s) &=& 2\rho(s) T^I_{J,2}(s){\rm Re}{T^I_{J,4}(s)}\ ,\\ %
...... \nonumber
\eqa%
With these relations it is easy to prove that the [1,2] Pad\'e
approximant in Eq.(\ref{pade}) satisfies elastic unitarity:%
\be%
{\rm Im}{{T^I_J}^{[1,2]}(s)} = \rho(s)|{T^I_J}^{[1,2]}(s)|^2\ .%
\ee%
In the following we frequently omit the indices $I,J$ of the $T$
matrix for simplicity if it causes no confusion. For any given
amplitude satisfying single channel unitarity, following the
method of Refs.~\cite{Xiao01,Xiao03}, we define two real analytic
functions $\tilde{F}$ and $F$ as%
\bqa\label{FtF}
\tilde{F}(s) &=& {1\over 2}{\left(S(s)+{1\over S(s)}\right)}\ ,\nonumber \\%
F(s) &=& {1\over {2i\rho(s)}}{\left(S(s)-{1\over S(s)}\right)}\ .%
\eqa%
It is obvious that $\tilde{F}$ and $\rho F$ are the analytic
continuation of $\cos (2\delta)$ and $\sin (2\delta)$, as the
scattering $S$ matrix is equal to $exp\{2i\delta\}$ in the
physical region.  According to \cite{Xiao01,Xiao03},\ we have the
following dispersion relations for $F$ and $\tilde F$:
 \bqa\label{sin2d}%
 \sin(2\delta)&=&\rho F \ =\rho (
 \alpha+ \sum_i{\beta_i \over 2i\rho(s_i)(s-s_i)}-\sum_j{1\over
  2i\rho(z^{II}_j)S'(z^{II}_j)(s-z^{II}_j)}  \nonumber \\ &&
  +{1\over\pi}\int_L{{\rm Im}_LF(s')
  \over s'-s} ds')\ ,
  \eqa
  and,
   \bqa
   \cos(2\delta) &=&\tilde{F}
   =\tilde{\alpha}+\sum_i {\beta_i\over2(s-s_i)}
  +\sum_j{1\over
 2S'(z^{II}_j)(s-z^{II}_j)}\nonumber\\
  && +{1\over\pi}\int_L {{\rm
 Im}_L\tilde F(s')
 \over s'-s} ds' \ ,  \label{cos2d}
\eqa
 where $\alpha$ and $\tilde\alpha$ are subtraction constants,
 $s_i$ denotes the possible
bound state pole positions and $\beta_i$ denotes the corresponding
residues of $S$;  $z_j^{II}$ denotes either the possible resonance
pole positions on the second sheet, which are grouped into complex
conjugated pairs, or the virtual state pole positions when
$z_j^{II}$ is real. The integrals in Eqs.~(\ref{sin2d}) and
(\ref{cos2d}) denote the cut contributions and one subtraction to
each integral is understood, according to general physical
consideration.\footnote{In general the dispersion integrals in
Eqs.~(\ref{sin2d}) and (\ref{cos2d}) need one subtraction, but the
Pad\'e amplitude is special in that the integrals are finite and
need no subtraction. Therefore the subtraction constant, $\alpha$
and $\tilde \alpha$ are 0 and the integrals are unsubtracted when
analyzing the Pad\'e amplitude.} $L=(-\infty,0]$ is the left hand
cut ($l.h.c.$). The discontinuities on the left in
Eqs.~(\ref{sin2d}) and (\ref{cos2d}) satisfy the following
equations~\cite{Xiao03},
 \bqa
{\rm Im}_L\tilde{F}(s) &=& -2\rho (s){\rm Im}_L{\rm Im}_RT(s)\ ,\nonumber \\%
{\rm Im}_LF(s) &=& 2{\rm Im}_L{\rm Re}_RT(s)\ .%
\eqa%
 In order to evaluate the values of ${\rm Im}_L{\rm Re}_RT(s)$
and ${\rm Im}_L{\rm Im}_RT(s)$, we need the analytical expressions
of ${\rm Re}_RT(s)$ and ${\rm Im}_RT(s)$ which can be derived from
the expression of the $T^{[1,2]}(s)$ in Eq.~(\ref{pade}),
 \bqa\label{pp1}
{\rm Re}_RT^{[1,2]}(s) &=& {{T_2(s)^3a}\over {a^2+b^2}}\ ,\nonumber \\%
{\rm Im}_RT^{[1,2]}(s) &=& {{T_2(s)^3b}\over {a^2+b^2}}\ ,%
\eqa%
where
 \bqa\label{pp2} a &=&
T_2(s)^2-T_2(s)\mathrm{Re}_RT_4(s)-T_2(s)\mathrm{Re}_RT_6(s)
+(\mathrm{Re}_RT_4(s))^2-(\mathrm{Im}_RT_4(s))^2\ ,
\nonumber \\%
b &=&
T_2(s)\mathrm{Im}_RT_4(s)+T_2(s)\mathrm{Im}_RT_6(s)-2\mathrm{Re}_RT_4(s)\mathrm{Im}_RT_4(s)\ .%
\eqa%
Analytical expressions for ${\rm Im}_L\tilde F$ and ${\rm Im}_LF$
in terms of perturbation amplitudes can also be written down, or
can be calculated numerically from Eq.~(\ref{pp1}).
 In Eqs.~(\ref{sin2d}) and
(\ref{cos2d}) we did not include  resonance poles on the first
sheet, since they are not allowed physically. However, as we
stated before, the Pad\'e amplitude may contain SPSRs. When using
Eqs.~(\ref{sin2d}) and (\ref{cos2d}) to analyze the Pad\'e
amplitude the dispersion representations have to be modified to
include those terms representing SPSRs. This can easily be done by
using Eq.~(\ref{FtF}).

Making use of  the property of the scattering amplitude at
threshold one can recast Eqs.~(\ref{sin2d}) and (\ref{cos2d}) in
the following form:
 \bqa\label{sin2d'}%
 &&\sin(2\delta)=\rho(s) F \ =\rho(s) \{
 2a^I_J + \sum_i{\beta_i(4-s) \over 2i\rho(s_i)(s-s_i)(4-s_i)}\nonumber \\
 &&-\sum_j{4-s\over
  2i\rho(z^{II}_j)S'(z^{II}_j)(s-z^{II}_j)(4-z_j^{II})}
  +{s-4\over\pi}\int_L{{\rm Im}_LF(s')
  \over (s'-s)(s'-4)} ds'\}\ ,\nonumber\\
 &&  \cos(2\delta) =\tilde{F}
   =1+\sum_i {\beta_i(4-s)\over2(s-s_i)(4-s_i)}
  +\sum_j{4-s\over
 2S'(z^{II}_j)(s-z^{II}_j)(4-z_j^{II})}\nonumber\\
   &&+{s-4\over\pi}\int_L {{\rm
 Im}_L\tilde F(s')
 \over (s'-s)(s'-4)} ds' \ ,
\eqa
 in which $a_J^I$ represents the scattering length parameter in
 the corresponding channel. The difference between
 Eqs.~(\ref{sin2d}), (\ref{cos2d}) and Eq.~(\ref{sin2d'}) really
 makes the difference: in the latter formula a constant
 contribution is subtracted from each pole term. The new
 definition of the pole contribution (that is the original pole
 contribution minus the its contribution at $s=4$) only probes
 the $s$ dependence of the pole term. For example, in the limit
 $z_j^{II}\to \infty$ while $S'(z_j^{II})z_j^{II}$ is held fixed,
 the pole contributes a constant term to the dispersion relation
 according to Eqs.~(\ref{sin2d}) and (\ref{cos2d}). This constant
 term is reabsorbed into the scattering length parameter in
 Eq.~(\ref{sin2d'}) and the pole no longer contributes to the dispersion relation,
 according to the new definition of pole
 contribution.  Similar discussion can be made for the case of
 SPSR.
 In the following we will always use the new
 definition of the pole contribution.
Except the pole contributions, the rest of the $r.$$h.$$s.$ of
Eq.~(\ref{sin2d'}) will be called the background contribution in
the following text.

 The
Eq.~(\ref{sin2d'})  allows us to explicitly examine different
contributions from various kinds of dynamical singularities  to
the phase shifts. We have computed various contributions to
$\cos(2\delta)$ and $\sin(2\delta)$ in IJ=00,11 and 20 channels
both in [1,2] Pad\'e and [1,1] Pad\'e approximations as presented
below.

The $SU(2)\times SU(2)$  effective Lagrangian at $O(p^6)$ contains
two sets of parameters: $l_1$ -- $l_4$ of $O(p^4)$ and $r_1$ --
$r_6$ of $O(p^6)$. Here we take these parameters the same as in
Ref.~\cite{2loops}: the scale-independent couplings ${\bar l}_i$
 are,%
\be\label{constants-li}%
\bar l_1 = -1.7\ ,\ \ %
\bar l_2 = 6.1\ ,\ \ %
\bar l_3 = 2.9\ ,\ \ %
\bar l_4 = 4.3\ ;%
\ee%
and the constants $r^R_i$ (the resonance contributions to the
low-energy constants of $O(p^6)$) are,%
\bqa\label{constants-ri}%
&&r^R_1 = -0.6\times 10^{-4}\ ,\ \ %
r^R_2  = 1.3\times 10^{-4}\ ,\ \ %
r^R_3 = -1.7\times 10^{-4}\ ,\ \ \nonumber \\%
&&r^R_4 = -1.0\times 10^{-4}\ ,\ \ %
r^R_5 = 1.1\times 10^{-4}\ ,\ \ %
r^R_6 = 0.3\times 10^{-4}\ ,%
\eqa%
and we take the renormalization scale $\mu=1$GeV when evaluating
the constants $b_i$ appeared in Eq.~(\ref{A(s,t,u)}). Using the
above values of parameters we can determine poles and cuts of the
Pad\'e amplitudes. As shown in Table~\ref{table-2-loop}, Pad\'e
approximation not only predicts the existence of the $\sigma$ and
$\rho$ resonances,\ but also generates many other poles on the
complex $s$ plane, and more poles exist in [1,2] than in [1,1] Pad\'e approximant.%
\begin{table}[bt]%
\centering\vspace{-0.cm}%
\begin{tabular}{|c|c|c|c|c|}%
\hline%
IJ & poles & Re[$s_{p}$] & Im[$s_{p}$] & Res[$\tilde F$]  \\%
\hline%
00 & $\sigma$ & 457MeV(M) & 475MeV($\Gamma$) & --6.43--7.31i \\%
\cline{2-5}
   & R & 395MeV(M) & 2.17GeV($\Gamma$) & --29.12+16.19i  \\%
\cline{2-5}%
   & SPSR & --26.03 & 1.48 & 4.77--12.66i \\%
\cline{2-5}%
   & SPSR & 86.47 & 76.90 & 31.44+44.38i  \\%
\hline%
11 & $\rho$ & 648MeV(M) & 118MeV($\Gamma$) & --2.25+3.01i  \\%
\cline{2-5}%
   & SPSR & 69.25 & 19.22 & --0.14+17.89i  \\%
\hline%
20 & VS & 0.0513651 & 0 & 0.0477297  \\%
\cline{2-5}%
   & SPSR & --18.58 & 14.15 & 8.44+6.13i  \\%
\cline{2-5}%
   & SPSR & 135.14 & 32.16 & 0.38+32.34i  \\%
\cline{2-5}%
\hline%
\end{tabular}%
\caption{\label{table-2-loop}Resonances(R), spurious physical
sheet resonances (SPSR) and virtual states (VS) as predicted by
the [1,2] Pad\'e approximation on the complex $s$ plane using the
coupling constants given by Eq.~(\ref{constants-li}) and
Eq.~(\ref{constants-ri}). The pole
position $s_{p}=(M+i\Gamma/2)^2$. All numbers are in unit of $m_\pi^2$ unless otherwise stated.
The  values of the scattering length parameters are, $a^0_0=0.224, a^2_0=-0.0412$.}%
\end{table}%
\ In addition to the poles found in table~\ref{table-2-loop}, it
is found that there also exist 2 pairs of BS/VS poles located
close to the Adler Zero position of $T^I_{J,2}$ in both IJ=00 and
IJ=20 channels (in a wide range of the $\bar l_i$ and $r_i$
parameters). Similar to what happens in the [1,1] Pad\'e
case~\cite{ang} they can be tuned away within reasonable range of
the $\bar l_i$ and $r_i$ parameters and are only artifacts of the
Pad\'e approximants. More importantly they only have very tiny
effects and can be safely neglected. The existence of the virtual
state in the IJ=20 channel has been clarified in Ref.~\cite{ang}
but its effect is also very small.

Besides those well established resonances which can be found in
table~1, Pad\'e approximants predict resonances or SPSRs at
distant places on the complex $s$ plane. Chiral perturbation
expansion only works in a region close to the threshold or
$|s|<<1$GeV$^2$, therefore the Pad\'e approximants should also be
expected to be reasonable only in a limited region:
$|z|<<1$GeV$^2$ on the complex $s$ plane. Hence any prediction
from Pad\'e approximants at distant places should not be
trustworthy, no matter the predicted poles are on  the first sheet
or on the second sheet. Of course we should not take these
predictions seriously. The real problem for the Pad\'e amplitude
is that in many cases the distant poles do not truly decouple from
the low energy physics, as indicated by their large couplings. The
Eqs.~(\ref{sin2d}) and (\ref{cos2d}) afford us a useful tool to
evaluate the influence of these spurious or unreliable
contributions quantitatively.

To have a clear insight to the problem we are facing we  perform
the following calculation: We use the MINUIT program to make a
global fit to the experimental phase shift in both the IJ=00,20
and 11 channels using the Pad\'e amplitudes. The
data~(\cite{rho}--\cite{E865}) are taken from the threshold to
730MeV in IJ=00 channel and to 1GeV in IJ=11 and 20 channels. The
fit results of the coupling constants $\bar l _i$ and $r_i$ are
listed in the following,
\bqa\label{constants-fit}%
&&\ \bar l_1 = -4.47^{+0.82}_{-0.83}\ ,\ \ \ %
\bar l_2 = 4.37^{+0.11}_{-0.11}\ ,\nonumber \\ %
&&\ \bar l_3 = -0.21^{+14.00}_{-13.71}\ ,\ \ %
\bar l_4 = 7.35^{+0.51}_{-0.52}\ ;\nonumber \\ %
&& r_1 = -84.85^{+363.88}_{-282.82}\times 10^{-4}\ ,\ \ %
r_2  = 3.96^{+11.99}_{-12.23}\times 10^{-4}\ ,\nonumber \\ %
&& r_3 = -34.36^{+5.34}_{-5.88}\times 10^{-4}\ ,\ \ \ \ %
r_4 = -0.11^{+1.24}_{-1.23}\times 10^{-4}\ ,\nonumber \\ %
&& r_5 = 7.66^{+1.24}_{-1.13}\times 10^{-4}\ ,\ \ \ \ \ \ \ \ %
r_6 = -2.61^{+0.25}_{-0.26}\times 10^{-4}\ .%
\eqa%
From above results we find that the $\chi^2$ fit is very
insensitive to $\bar l_3$, $r_1$ and $r_2$. The other $r_i$
parameters are barely comparable in order of magnitude to those
given in Eq.~(\ref{constants-ri})~\cite{fit}. The results for the
fit are shown in table~\ref{table-2-loop-fit} and
Figs.~\ref{delta00-cos-fit}--\ref{delta20-cos-fit}. In order to
compare the [1,2] Pad\'e results with [1,1] Pad\'e results we also
made the global fit in the latter case. The results are given in
table~\ref{table-1-loop}. In Fig.~\ref{fit1loop} we plot the [1,1]
Pad\'e fit results in the most interesting IJ=00 channel. In
Figs.~1 -- 4, the solid lines represent the phase shift value
directly obtained from the Pad\'e amplitudes in the physical
region, $i.e.$, $\delta_\pi(s)={\rm Atan}({{\rm Im}_RT(s)}/{{\rm
Re}_RT(s)})$. The dash-dot-dot-dash lines in Figs.~1--3 and the
dash-dot-dash line in Fig.~4 represent the summation of all
contributions on the $r.h.s.$ of Eqs.~(\ref{sin2d}) and
(\ref{cos2d}) (also including SPSRs' contributions). The two kinds
of lines must coincide with each other as a consistency check of
our numerical calculation.
\begin{table}[bt]%
\centering\vspace{-0.cm}%
\begin{tabular}{|c|c|c|c|c|}%
\hline%
IJ & Pole & Re[$s_{p}$] & Im[$s_{p}$] & Res$\tilde F$  \\%
\hline%
00 & $\sigma$ & 586MeV(M) & 775MeV($\Gamma$) & --42.28+2.22i \\%
\cline{2-5}
   & R & 318MeV(M) & 935MeV($\Gamma$) & 18.28+15.41i  \\%
\cline{2-5}%
   & SPSR & 135.68 & 56.46 & 22.82+39.79i  \\%
\cline{2-5}%
\hline%
11 & $\rho$ & 768MeV(M) & 150MeV($\Gamma$) & --1.55+6.13i  \\%
\cline{2-5}%
   & R & 678MeV(M) & 1.02GeV($\Gamma$) & 5.65--2.24i \\%
\cline{2-5}%
   & SPSR & 0.83 & 39.17 & --4.13+1.28i  \\%
\hline%
20 & VS & 0.0483292 & 0 & 0.0451211  \\%
\cline{2-5}
   & R & 647MeV(M) & 2.34GeV($\Gamma$) & --31.05+41.92i  \\%
\cline{2-5}
   & R & 299MeV(M) & 7.27GeV($\Gamma$) & 96.25--29.84i  \\%
\cline{2-5}%
   & SPSR & --11.94 & 27.44 & 9.62--13.06i \\%
\cline{2-5}%
\hline%
\end{tabular}%
\caption{\label{table-2-loop-fit}Resonances(R), spurious physical
sheet resonances (SPSR) and virtual states (VS) as predicted by
the [1,2] Pad\'e approximation on the complex $s$ plane using the
values from Eq.~(\ref{constants-fit}). The pole position
$s_{p}=(M+i\Gamma/2)^2$. All numbers in the table are in unit of $m_\pi^2$ unless stated otherwise
The fit values of the scattering length parameters are, $a^0_0=0.217, a^2_0=-0.0531$.}%
\end{table}%
The coupling constants obtained from the fit using the [1,1]
Pad\'e amplitudes are the following,
 \bqa\label{r1loop}
\bar l_1 = -0.33^{+0.067}_{-0.068}\ ,\ %
\bar l_2 = 5.83^{+0.067}_{-0.067}\ ,\ %
\bar l_3 = 24.09^{+11.24}_{-10.52}\ ,\  %
\bar l_4 = 3.51^{+0.27}_{-0.27}\ . %
\eqa%
\begin{table}[bt]%
\centering\vspace{-0.cm}%
\begin{tabular}{|c|c|c|c|c|}%
\hline%
IJ & Pole & Re[$s_p$] & Im[$s_p$] & Res[$\tilde F$]  \\%
\hline%
00 & $\sigma$ & 456MeV(M) & 463MeV($\Gamma$) & --4.77--6.47i \\%
\cline{2-5}%
   & SPSR & --74.50 & 53.35 & --106.31--103.32i  \\%
\cline{2-5}
\hline%
11 & $\rho$ & 751MeV(M) & 144MeV($\Gamma$) & --2.55+4.99i \\%
\hline%
20 & VS & 0.0335461 & 0 & 0.0318398  \\%
\cline{2-5}%
   & SPSR & 103.30 & 351.11 & --489.54+77.33i  \\%
\cline{2-5}%
\hline%
\end{tabular}%
\caption{\label{table-1-loop}Resonances(R), spurious physical
sheet resonances (SPSR) and virtual states (VS) as predicted by
the [1,1] Pad\'e approximation on the complex $s$ plane using the
values from Eq.~(\ref{constants-li}) and Eq.(\ref{constants-ri}).
The pole position $s_{pole}=(M+i\Gamma/2)^2$. All numbers in the
table are in unit of $m_\pi^2$ unless stated otherwise.
The fit values of the scattering length parameters are, $a^0_0=0.186, a^2_0=-0.0467$.}%
\end{table}%

Comparing the results of [1,1] Pad\'e approximants and the [1,2]
Pad\'e approximants we conclude that the [1,2] Pad\'e
approximation  does not in general improve the bad situation the
[1,1] Pad\'e approximants encounter, though the former can give a
better fit to the phase shift data\footnote{This is hardly
surprising since it contains 6 more parameters.}. In the IJ=11
channel the [1,1] Pad\'e amplitude only predict the $\rho$
resonance, in the [1,2] amplitude there are additional poles at
distant positions but their contributions are rather small, so
both the two amplitudes can be considered as phenomenologically
successful. In the IJ=20 channel both amplitudes run into disaster
since both of them predict huge contributions from SPSRs.
Comparing Fig.~\ref{delta11-cos-fit} with Fig.~2 of
Ref.~\cite{ang} one finds that the [1,2] Pad\'e amplitude is even
worse in the sense that it predicts large SPSR contribution also
to $\sin(2\delta)$. In fact the perturbation result in this
channel is much better~\cite{Xiao01}. In the most interesting
IJ=00 channel, the situation is more complicated. From
Fig.~\ref{fit1loop} we find that the SPSR's contribution to
$\sin(2\delta$) is sizable yet in Fig.~\ref{delta00-cos-fit} the
SPSR's contribution is very small. However, in the latter case
there appears another large contribution from a resonance ($R$)
located at $s_R=(0.318\pm 0.468i)^2$GeV$^2$ in the IJ=00 channel,
which is neither very far away from nor very close to the low
energy physics region. When looking at
table~\ref{table-2-loop-fit} one may even confuse the two
resonances  in the IJ=00 channel: the $\sigma$ and $R$. The two
resonances are distinguishable in the following way: in
table~\ref{table-2-loop} one resonance pole is much closer to the
physical region comparing with another one, the former is of
course denoted as $\sigma$.~\footnote{The question whether this
resonance is the $\sigma$ meson responsible for spontaneous chiral
symmetry breaking is another question, for recent discussions, see
for example Refs.~\cite{sigma1} and \cite{sigma2}.} Then we can
tune the parameters $\bar l_i$ and $r_i$ continuously, from the
values given in Eqs.~(\ref{constants-li}) and (\ref{constants-ri})
to the central
 values of the parameters given in Eq.~(\ref{constants-fit}), and
 keep track of the pole positions. In this way we can distinguish
the two resonance poles in table~\ref{table-2-loop-fit}. Obviously
the pole position of $R$ is very sensitive to the parameters of
the chiral Lagrangian and the pole position of the $\sigma$ meson
is rather stable. The former comes actually from very distant
places, therefore its existence and contribution is still doubtful
even though it is located not very far from the physical region as
predicted by the parameter set determined from the global fit.
 In this sense one hesitates to conclude that
the [1,2] Pad\'e approximant in the IJ=00 channel is better than
the [1,1] approximation even though the SPSRs' contribution is
reduced.

In above discussions we point out and study in details the problem
of the Pad\'e approximation method induced by the existence of
SPSRs, which has not drawn much attention in the literature. It is
worth mentioning that the author of Ref.~\cite{hannah} noticed
such a problem and proposed a scheme to remedy the situation when
studying the $\pi$ scalar form-factor, $F$, to two loops using
IAM.\footnote{Similar discussion on $\pi\pi$ scattering amplitudes
is not yet found in the literature.}  When obtaining Eq.~(7) from
Eq.~(6) in that paper, the author essentially did the following:
expanding the SPSR term up to $O(s^2)$ and neglecting the higher
order terms, and reabsorb the coefficients of the second order
polynomial into the low energy constants, and the latter are
determined by either CHPT or by fit. In this way the SPSR (which,
according to the author, locates on the negative real axis) is
eliminated in the newly obtained $F$. This effort is respectable
as it is in the right direction, but is still problematic.
Actually it is not difficult to prove that the procedure proposed
by the author does not make the effects of the spurious pole
totally vanish but actually moves the pole from the 1st sheet to
the 2nd sheet (see the appendix for the proof).\footnote{It is
worth pointing out that the author wrongfully claimed that the IAM
amplitude contains infinitely many sheets which actually contains
only two.} A pole on the negative real axis on the 2nd sheet is
still dubious, even though its numerical influence is not
estimated.

To conclude, in general the prediction from the [1,2] Pad\'e
approximation is not in any sense more trustworthy than the [1,1]
Pad\'e approximation. A lesson one may draw from the discussion
made in this paper is that physics at distant places (no matter
spurious or not) as predicted by the Pad\'e approximation do not
necessarily decouple at low energies. We suggest, to make safe use
of the Pad\'e approximation one has to make a case by case
analysis to the amplitudes using the method proposed in this
paper. The smallness of the contribution from high energies may be
considered as a necessary condition for the predictions of the
Pad\'e amplitude at moderately low energies to be numerically
trustworthy.

This work is supported in part by China National Natural Science
Foundation under grant number 10047003 and 10055003.

\begin{newpage}
\appendix
\section{Appendix}
The proof follows: denote the formfactor defined in Eq.~(6) in
Ref.~\cite{hannah} by $F_1$ and that defined in Eq.~(7) by $F$.
The major difference between $F$ and $F_1$ is that the latter
contains the SPSR. Especially we have {\rm Im}$F$={\rm Im}$F_1$
and
\begin{equation}\label{1}
F= P_2(s)+{s^3\over\pi}\int_{4m_\pi^2}^\infty ds'\frac{{\rm
Im}F(s')}{s'^3(s'-s-i\epsilon)}\equiv {\rm Re}F(s)+i{\rm Im}F(s)\
.
\end{equation}
This $F$, even though does not satisfy exact unitarity, maintains
the two--sheet structure and can be analytically continued to the
second sheet by changing the sign of the kinematic factor
$\rho=\sqrt{1-4m_\pi^2/ s}$, notice that both ${\rm Re}F(s)$ and
$i{\rm Im}F(s)$ can be analytically continued to the complex $s$
plane. To be specific $F^{II}$ is,
\begin{equation}
F^{II}= P_2(s)+{s^3\over\pi}P\int_{4m_\pi^2}^\infty ds'\frac{{\rm
Im}F(s')}{s'^3(s'-s)}- i{\rm Im}F(s)= {\rm Re}F(s)-i{\rm Im}F(s)\
,
\end{equation}
$i.e.$, the $F^{II}$  is obtainable by just simply changing the
sign of ${\rm Im}F(s)$ and $P$ stands for the principle value. By
construction $F$ does not contain any poles on the entire cut
plane and according to Ref.~\cite{hannah} $F^{II}$ does. That is
possible only when both ${\rm Re}F(s)$ and $i{\rm Im}F(s)$ contain
poles on the cut plane but cancel each other in ${\rm
Re}F(s)+i{\rm Im}F(s)$, but not in ${\rm Re}F(s)-i{\rm Im}F(s)$.
Therefore the pole locations of $F^{II}$ are the same as $i{\rm
Im}F(s)$ and are also the same as $i{\rm Im}F_1(s)$. However it is
clear from  Eq.~(6) of Ref.~\cite{hannah} that $i{\rm Im}F_1(s)$
contains the poles of $F_1$ on the 1st sheet (corresponding to
zeros of $\Gamma^{(2)}$, for the definition of the latter see
Ref.~\cite{hannah}), as well as poles on the 2nd sheet
(corresponding to zeros of $\Gamma^{(2)*})$. Therefore all poles
of $F_1$ are transmitted into $F^{II}$ including both the $\sigma$
pole and the spurious pole.

\end{newpage}
\setlength{\topmargin}{5mm} \setlength{\textheight}{230mm}
\setlength{\textwidth}{150mm}
\begin{newpage}

\begin{figure}[htb]
\begin{center}
\mbox{\epsfxsize=120mm \epsffile{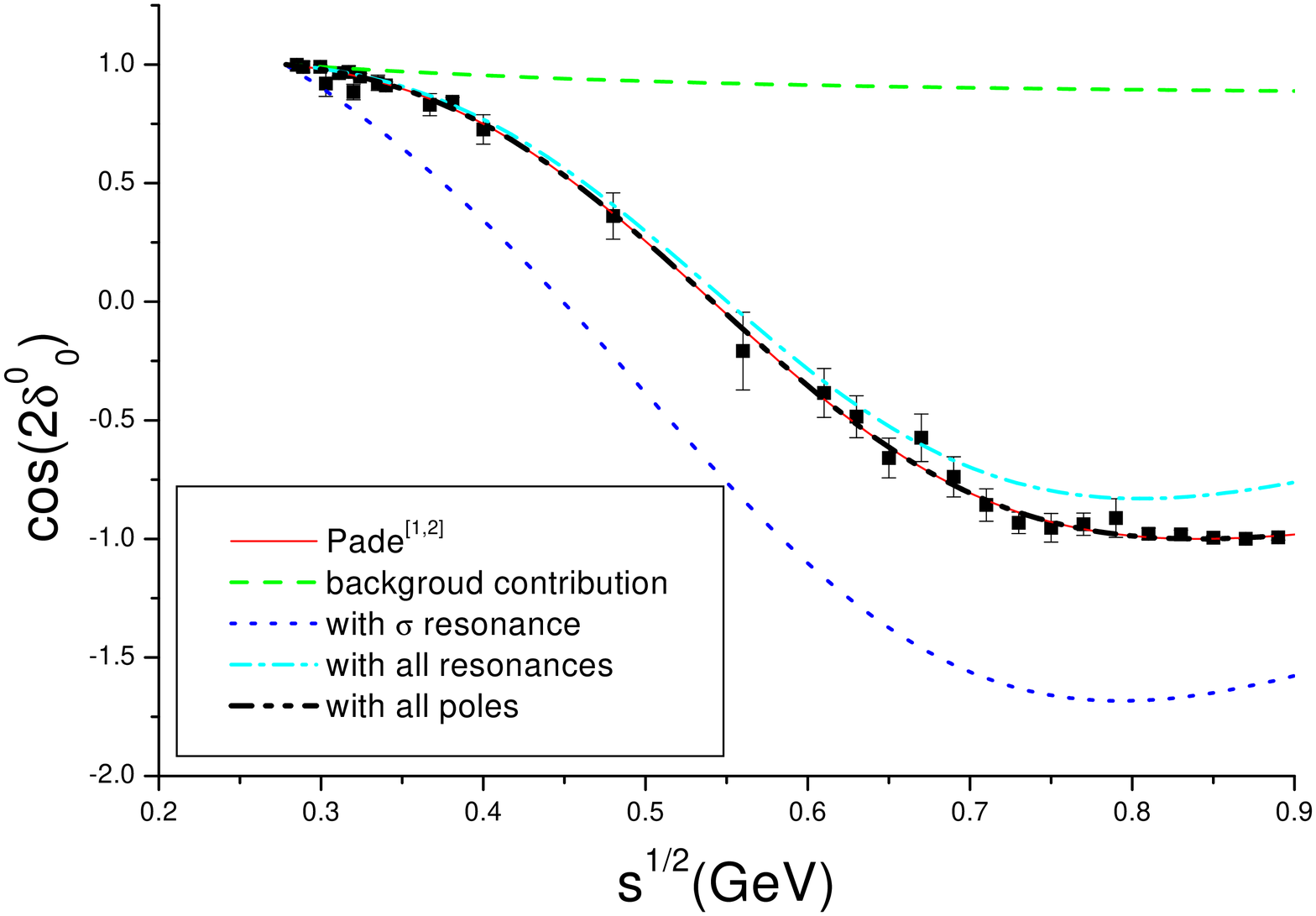}} \vspace*{5mm}%
\mbox{\epsfxsize=120mm \epsffile{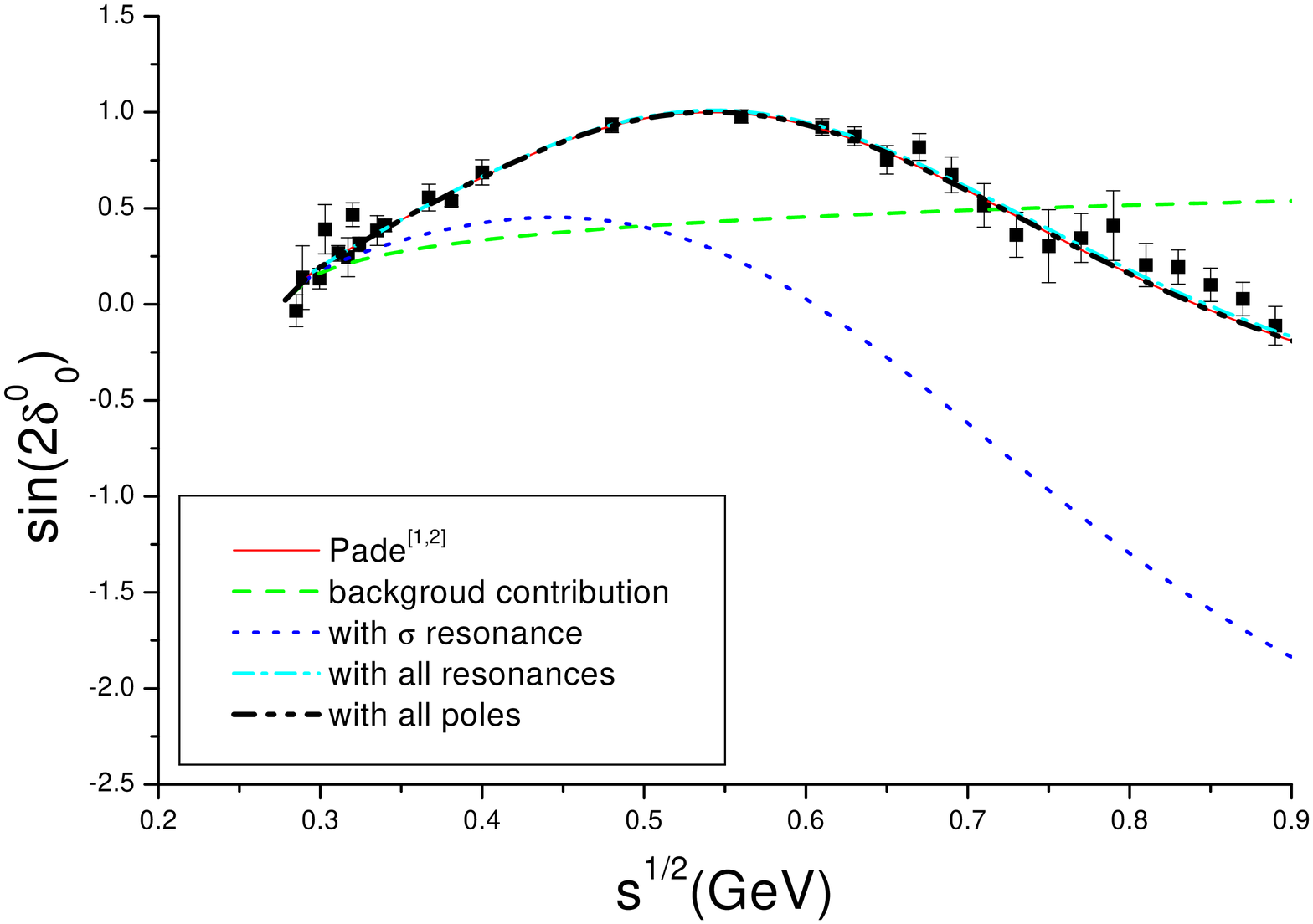}} \vspace*{0mm}%
\caption{ \label{delta00-cos-fit}Various contributions to
$\cos(2\delta^0_ 0)$ and $\sin(2\delta^0_ 0)$ in the IJ=00 channel
using the values from Eq.(\ref{constants-fit}). Notice that in the
IJ=00 channel we only fit the data up to 730MeV. The [1,2] Pad\'e
amplitude gives a much better description to the data above 730MeV
than the [1,1] Pad\'e amplitude, see Fig.~\ref{fit1loop} for
comparison. The solid line and the dash-dot-dot-dash line are
explained in the text. The dashed line is from the background
contribution only, whereas the dotted line represents the
contributions from the $\sigma$ and the background, and the
dash-dot-dash line is to add the resonance (R) contribution to the
dotted line. The fact that the dash-dot-dash line is very close to the solid line indicates
that the SPSR's contribution is very small.}%
\end{center}
\end{figure}
\end{newpage}
\begin{newpage}
\begin{figure}[htb]
\begin{center}
\mbox{\epsfxsize=120mm \epsffile{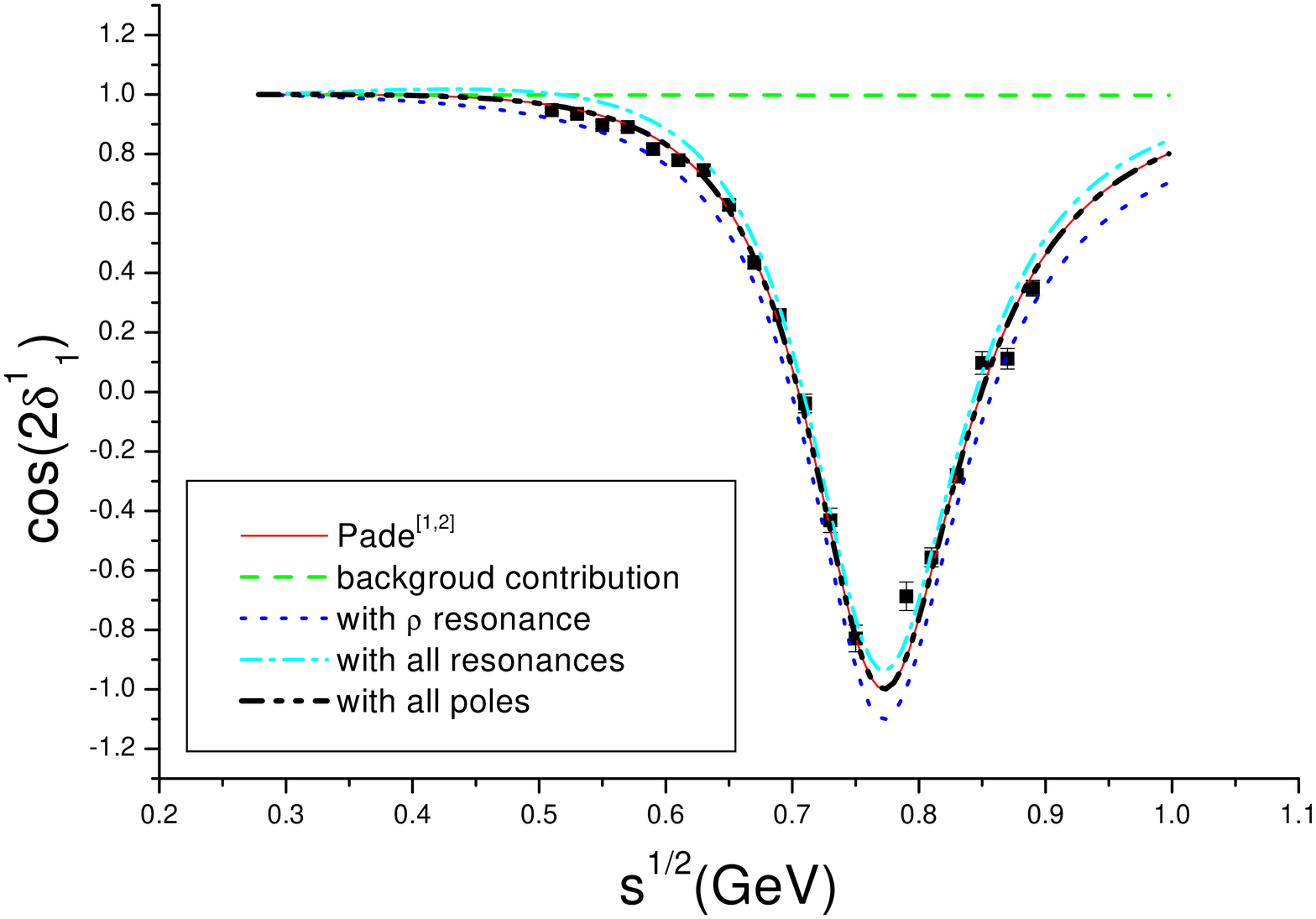}} \vspace*{5mm}%
\mbox{\epsfxsize=120mm \epsffile{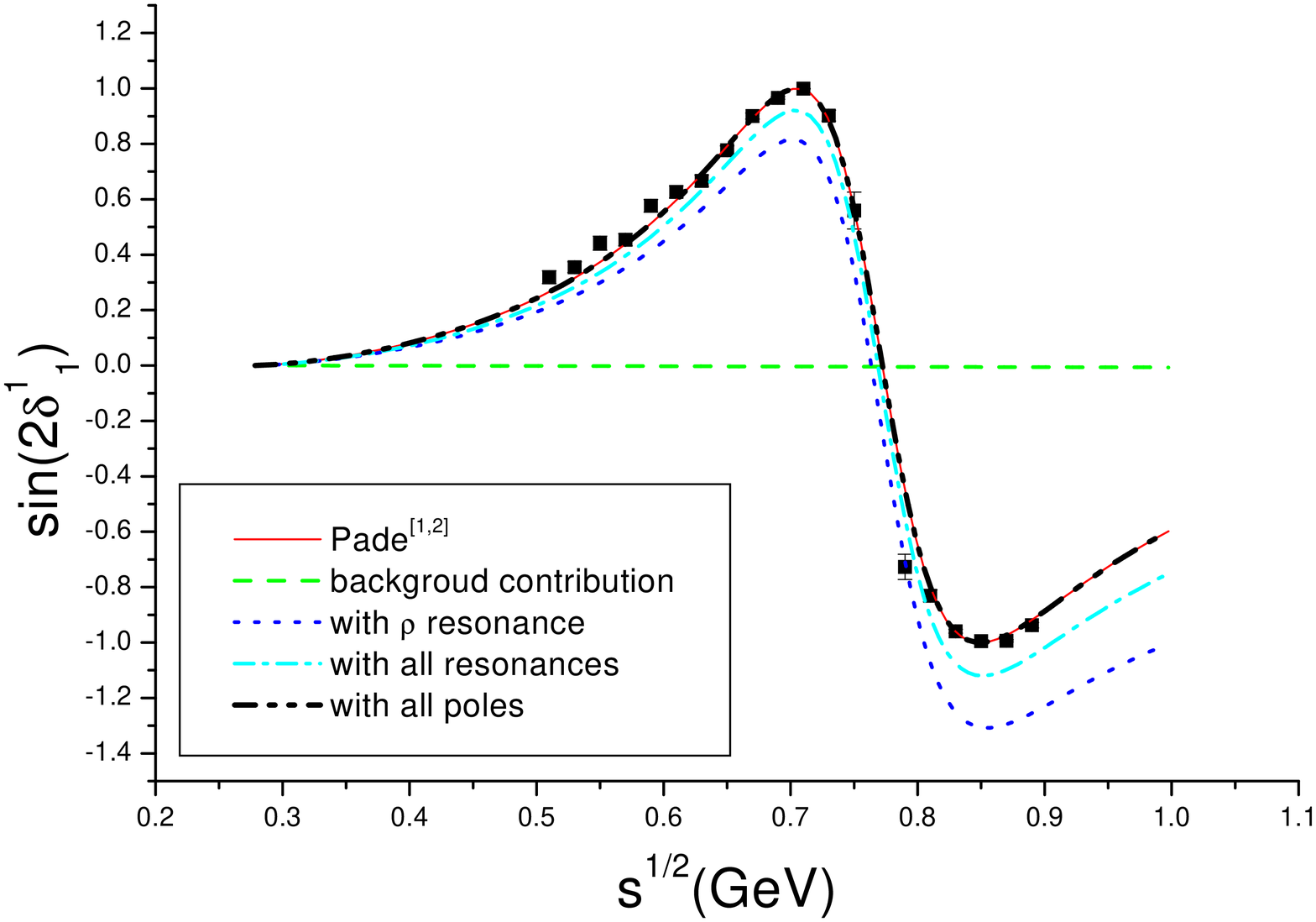}} \vspace*{0mm}%
\caption{ \label{delta11-cos-fit}Various contributions to
$\cos(2\delta^1_ 1)$ and $\sin(2\delta^1_ 1)$ in the IJ=11 channel
using the values from Eq.(\ref{constants-fit}). The meaning of
different lines is similar to those in Fig.~1.
}%
\end{center}
\end{figure}
\end{newpage}
\begin{newpage}
\begin{figure}[htb]
\begin{center}
\mbox{\epsfxsize=120mm \epsffile{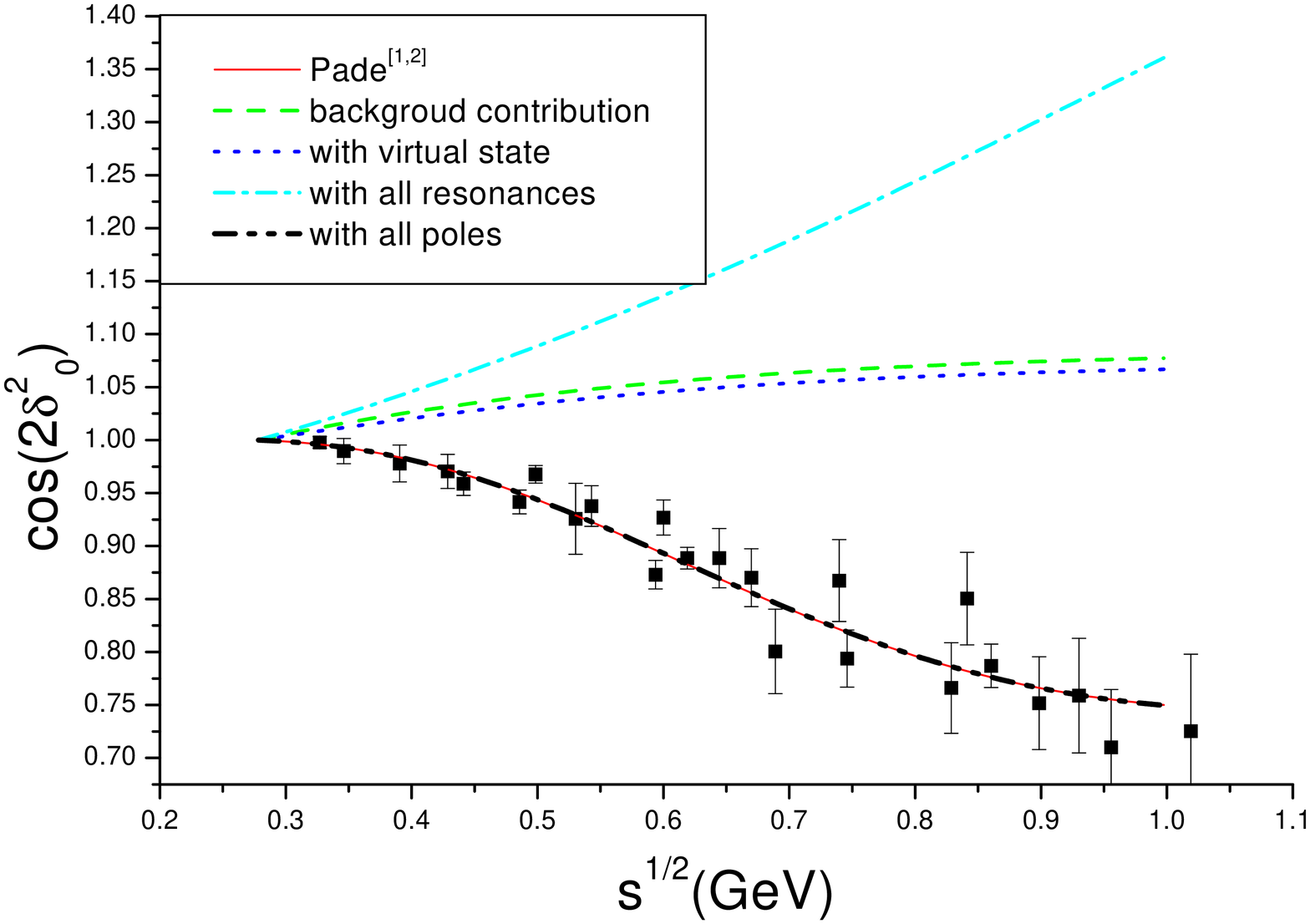}} \vspace*{5mm}%
\mbox{\epsfxsize=120mm \epsffile{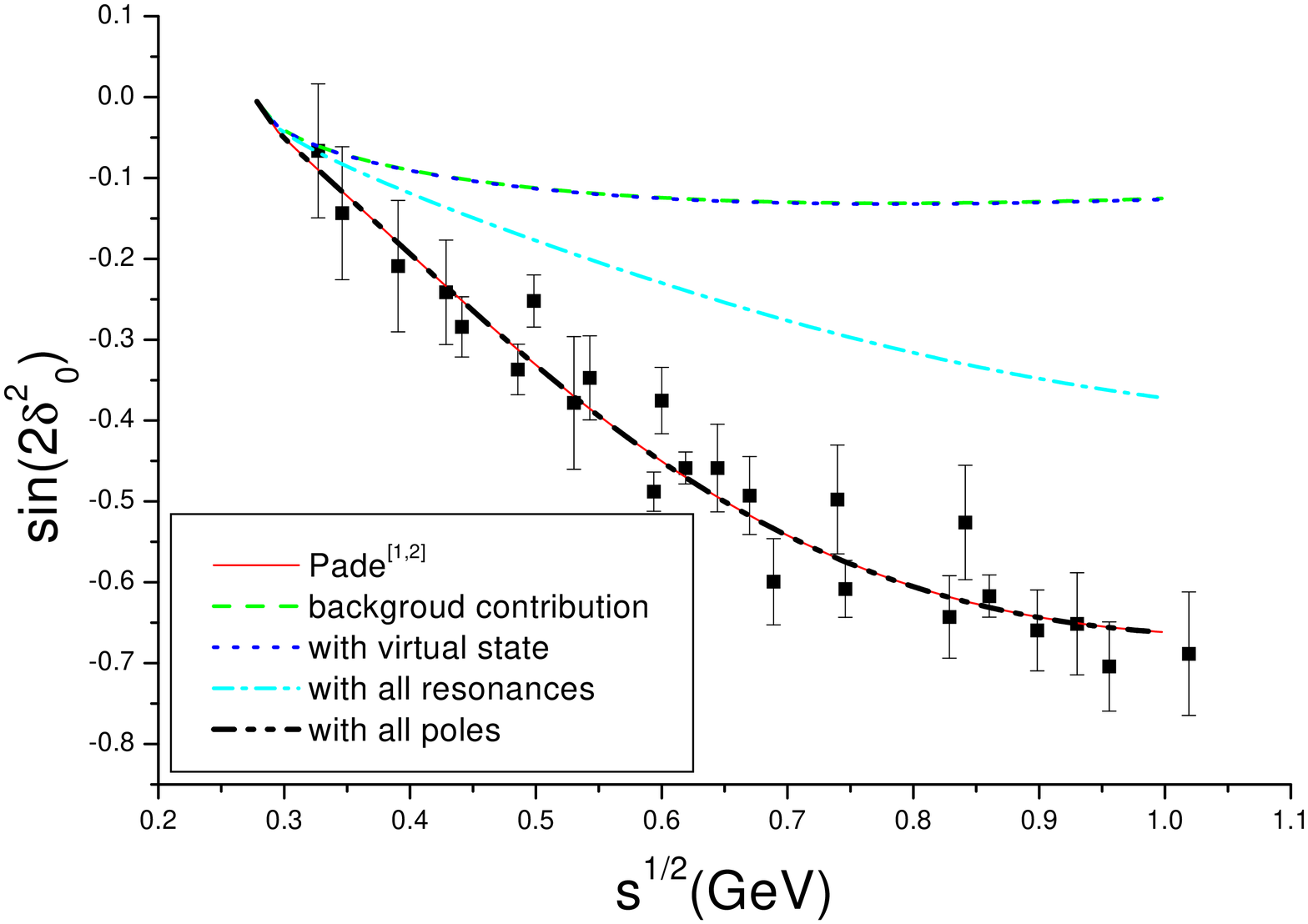}} \vspace*{0mm}%
\caption{ \label{delta20-cos-fit}Various contributions to
$\cos(2\delta^2_ 0)$ and $\sin(2\delta^2_ 0)$ in the IJ=20 channel
using the values from Eq.~(\ref{constants-fit}). The virtual state contribution
as represented by the difference between the dashed line and the dotted line is very small.}%
\end{center}
\end{figure}
\end{newpage}
\begin{newpage}
\begin{figure}[htb]
\begin{center}
\mbox{\epsfxsize=120mm \epsffile{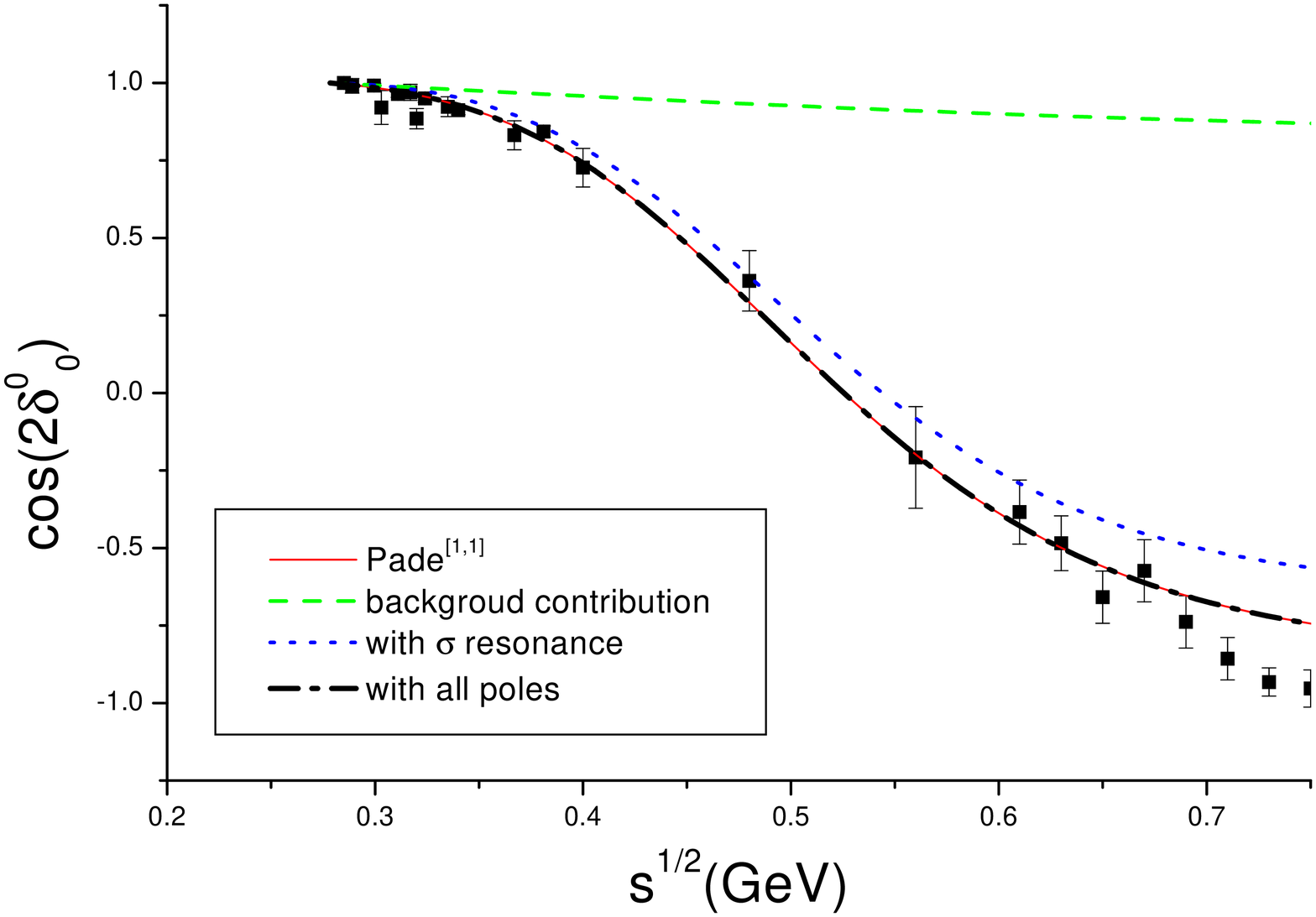}} \vspace*{5mm}%
\mbox{\epsfxsize=120mm \epsffile{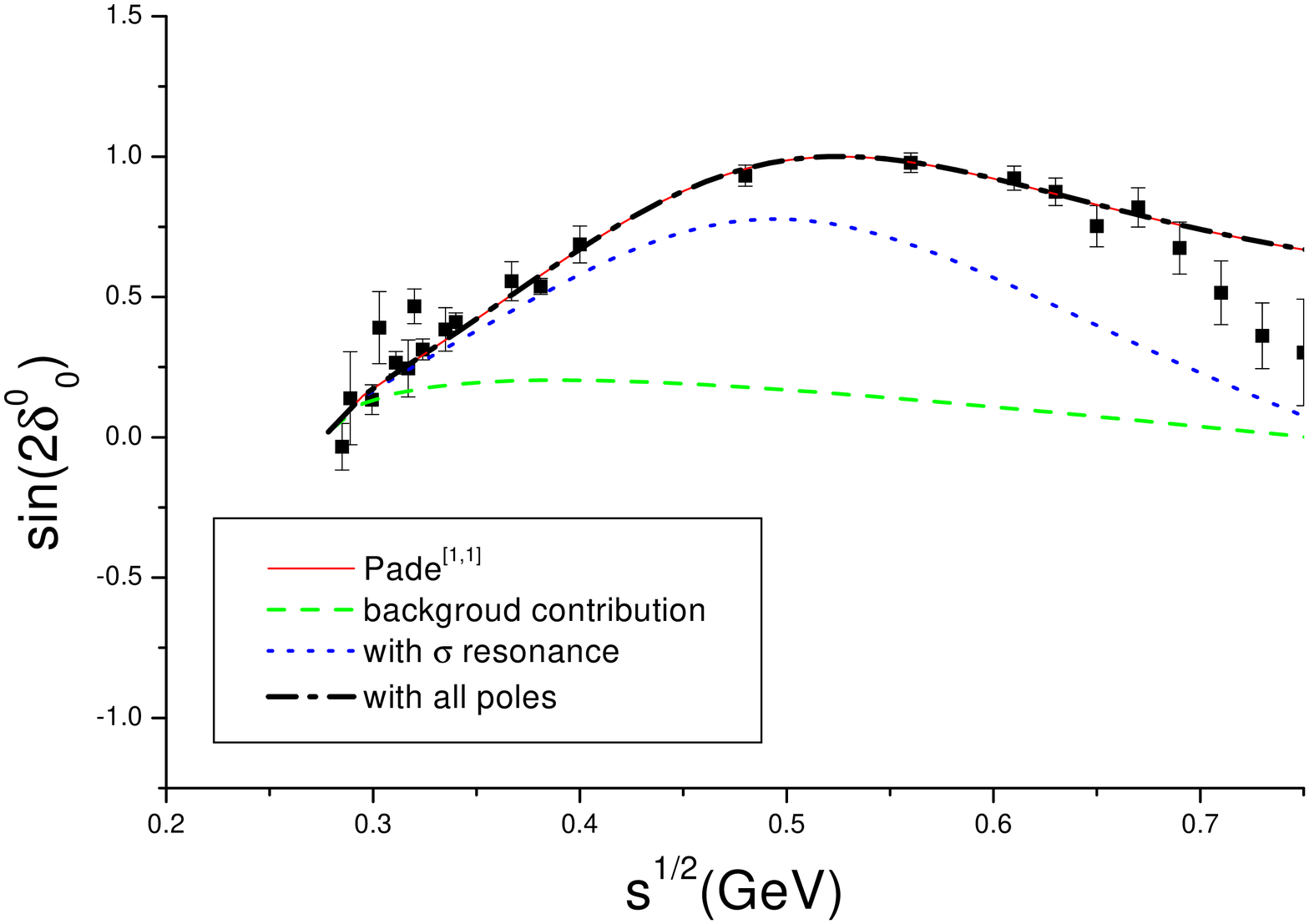}} \vspace*{0mm}%
\caption{ \label{fit1loop}Various contributions to
$\cos(2\delta^0_ 0)$ and $\sin(2\delta^0_ 0)$ in the IJ=00 channel
using the values from Eq.~(\ref{r1loop}), using the [1,1] Pad\'e amplitude.}%
\end{center}
\end{figure}
\end{newpage}

\end{document}